# Advancing Multiscale Structural Mapping for Alzheimer's Disease using Local Gyrification Index

Authors:

Jinhee Jang[1†], Geonwoo Baek[2†], Ikbeom Jang[2]*

Affiliations:

[1] Seoul St. Mary's Hospital, College of Medicine, The Catholic University of Korea, Seoul, South Korea

[2] Division of Computer Engineering, Hankuk University of Foreign Studies, Yongin, South Korea

Presenting author:

Full name: Jinhee Jang

Affiliation: Seoul St. Mary's Hospital, College of Medicine, The Catholic University of Korea, Seoul, South Korea

Email: znee@catholic.ac.kr

Key information:

Research question: This study aims to find whether other neurostructural measurements could be added and combined with the state-of-the-art Alzheimer's imaging marker called MSSM to improve sensitivity to neurodegeneration in Alzheimer's disease patients.

Findings: By applying various neurostructural measurements such as the local gyrification index and Jacobian white to the existing Multiscale Structural Mapping of Alzheimer's Disease Neurodegeneration, better results were obtained compared to previous methods, with the addition of LGI proving to be the most effective.

Meaning: The extended MSSM imaging marker may provide better ability for the detection of degeneration in Alzheimer's disease. This research shows that this method, using a single standard T1-weighted MRI, can support clinical diagnostics and help identify individuals who may need further biomarker evaluation.

†: equal contribution, *: corresponding author

# MANUSCRIPT

## Introduction

Early detection of Alzheimer's disease typically requires invasive tests such as positron emission tomography (PET) and lumbar puncture cerebrospinal fluid assessment. Recent studies have shown a possibility of detecting Alzheimer's disease neurodegeneration and neuropathology using a standard T1-weighted MRI (Jang et al. 2022, Jang et al. 2024). Early detection is crucial for timely treatment, highlighting the need for non-invasive methods. The proposed Multiscale Structural Mapping (MSSM) leverages MRI data to identify significant structural changes, highlighting differences between cognitively normal (CN) individuals and those with Alzheimer's disease (AD). MSSM combines the ratios of white matter signals at two depths and gray matter signals at four depths, resulting in eight values, along with cortical thickness, into a single value through dimensionality reduction. In this study, we aim to further improve the MSSM procedure via feature engineering. Important biomarkers used include vertex-wise cortical thickness, known indicators of AD. We iteratively added and combined various neurostructural measurement maps to MSSM to increase sensitivity to structural differences between the CN and AD groups.

## Material and methods

In this study, we utilized portions of the dataset obtained from phases ADNI 2 and ADNI 3 of the Alzheimer's Disease Neuroimaging Initiative (ADNI) project. The details of the MRI data and the participants are provided in Table 1. The study participants totaled 200. Participants were divided into two groups: Alzheimer's disease (AD) patients and cognitively normal (CN) individuals, with each group comprising 100 members. We categorized the candidates based on the inclusion of each vertex-wise data and the application of age correction. Alzheimer's and aging both cause structural changes such as cortical shrinkage; hence, controlling for age through age correction helps in distinguishing the effects due to AD more clearly. We evaluated the MSSM_Ext using the number of statistically significant vertices, with values representing $-\log_{10}(p^\wedge)$ that exceeded 1.3 (indicating a $p^\wedge$-value $< 0.05$), obtained from group analysis and Monte-Carlo simulation to correct the p-value using FreeSurfer.

## Results

T The MSSM_Ext significantly improved performance, achieving approximately three times the efficiency of using cortical thickness alone and about 1.5 times that of MSSM. This comparison was based on the sum of the percentages derived from the number of vertices with significant values $-\log_{10}(p^*) > 1.3$, excluding the corpus callosum. MSSM_Ext is MSSM combined with the Local Gyrification Index, the most beneficial feature for distinguishing between normal samples and Alzheimer's. The Local Gyrification Index (LGI) measures the complexity of the brain's surface, specifically the degree of folding of the cerebral cortex.

## Discussion and Conclusion

This study successfully improved MSSM by incorporating new data and applying feature engineering techniques like age correction. The results demonstrate the potential for further improvements in MSSM, which will be especially valuable when invasive methods are not feasible. Future applications will extend beyond current limitations as we refine our approach. The current method is limited to data that can be applied to T1-weighted MRI. In the future, we plan to expand MSSM_Ext to be applicable to a broader range of MRI data.

## Disclosures

There are no disclosures to report.

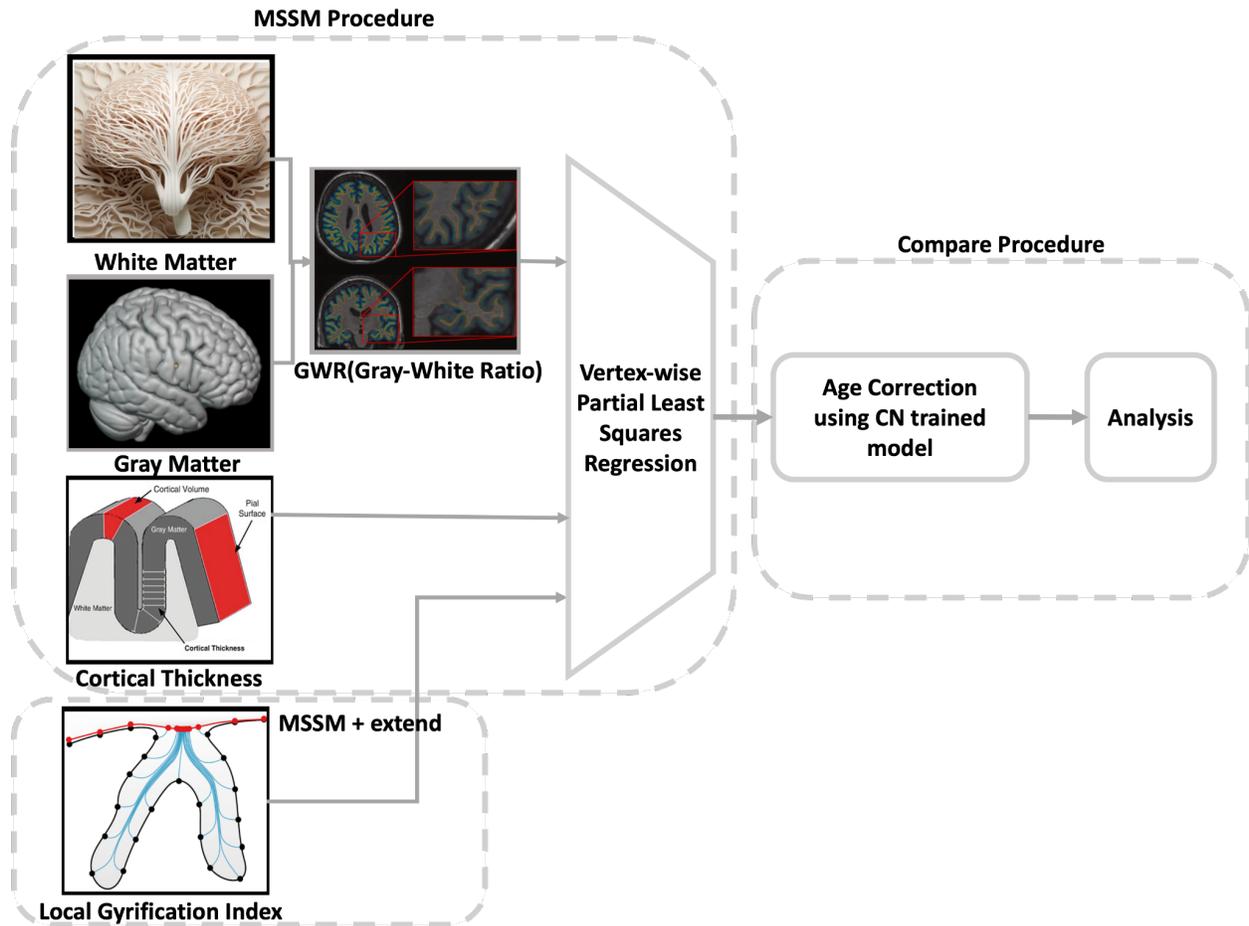

Figure 1: Diagram of the proposed MSSM_Ext. Normalization was applied to white matter, gray matter, cortical thickness, and the local gyrification index, followed by dimension reduction to a single feature from a total of 10 features (GWR, 8 maps, cortical thickness, local gyrification index) using partial least squares regression. Next, to remove the influence of age, age correction was performed using a model trained on the data of the cognitively normal group, and the residuals between the actual data and the model predictions were used. These residuals were used to perform group analysis by conducting GLM analysis to obtain the p-value, which was then corrected using Monte-Carlo simulation between the AD group and the CN group.

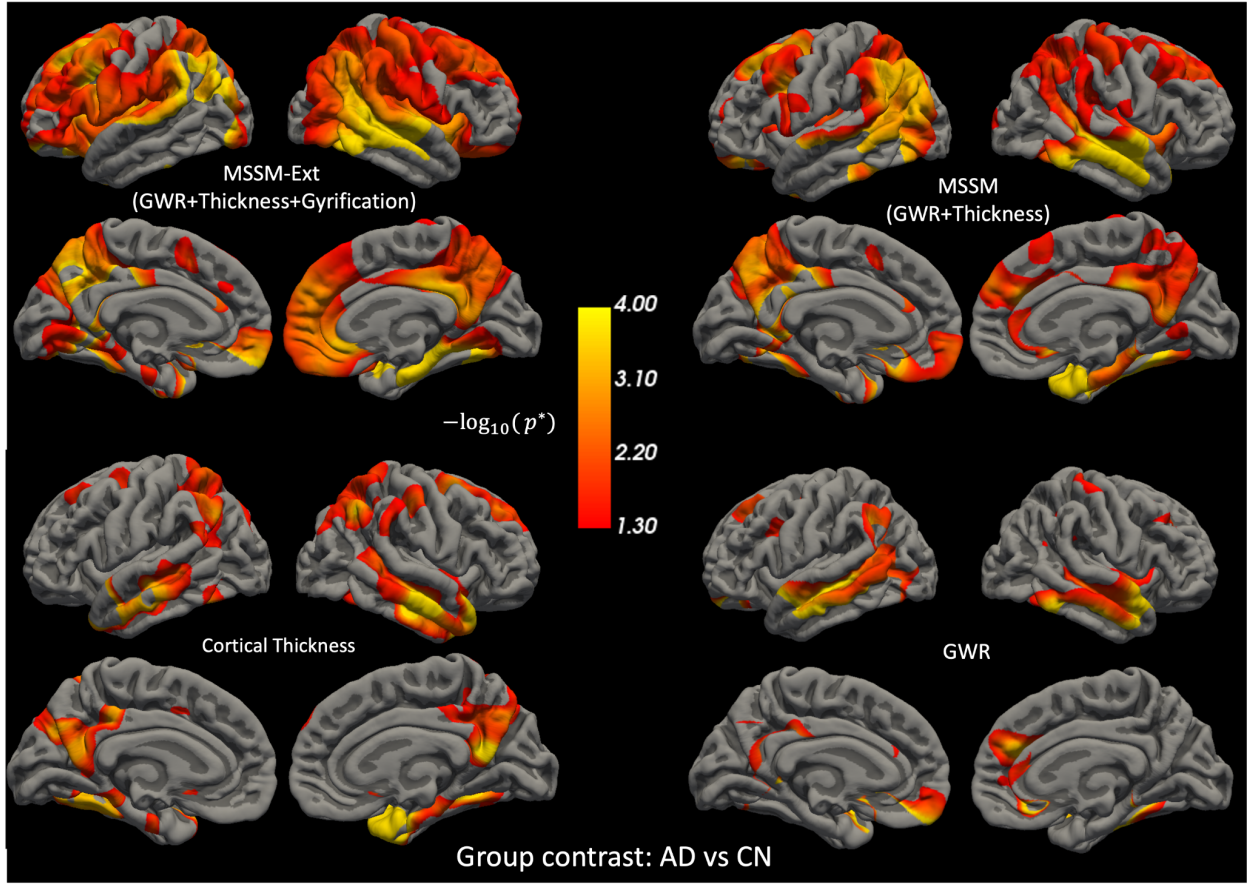

Figure 2: Effects of Alzheimer's disease (AD vs CN) on MSSM_Ext and comparison with other methods. The extent of the AD effect significantly increased compared to the original MSSM method, GWR, and cortical thickness. -log10(p*) is shown, where p* denotes the p-value in a t-test after multiple comparison correction.

|  | N | Age (Average) | Sex |
|---|---|---|---|
| AD | 100 | 75.86 | M:F = 57:43 |
| CN | 100 | 80.91 | M:F = 50:50 |

Table 1. The dataset used in this study consists of 100 individuals with Alzheimer's Disease (AD) and 100 cognitively normal (CN) individuals. The average age in the AD group is 75.86 years, with a gender ratio of 57 males to 43 females. The CN group has an average age of 80.91 years, with a gender ratio of 50 males to 50 females. The data was collected from ADNI (Alzheimer's Disease Neuroimaging Initiative) using the T1-weighted sagittal MPRAGE method on MRI systems from GE, Siemens, and Philips. The parameters used are TR = 2300 ms, TE = 2.98 ms, flip angle = 9, and voxel size = $1 \times 1 \times 1.2$ mm³.

|  | Significant vertices (left-hemi, %) | Significant vertices (right-hemi, %) | Significant vertices (both, %) |
| --- | --- | --- | --- |
| Cortical Thickness | 22.47 | 26.74 | 24.61 |
| GWR | 12.28 | 13.65 | 12.97 |
| MSSM (GWR+Thickness) | 41.48 | 44.57 | 43.03 |
| MSSM_Ext (GWR+Thickness +Local Gyrification Index) | 58.99 | 69.81 | 64.40 |

Table 2. Performance of the proposed method compared with other methods. Left/Right refers to the Left Hemisphere and Right Hemisphere. The Area Rate is the percentage calculated by dividing the number of vertices with (-log10(p*) > 1.3) by the number of vertices excluding the corpus callosum. The number of vertices excluding the corpus callosum is 149,955 for the left hemisphere and 149,926 for the right hemisphere.